# Investigation of AC Loss in HTS Cross-Conductor Cables for Electrical Power Transmission

Boyang Shen, T. A. Coombs, and Francesco Grilli

*Abstract* — This paper presents the alternating current (AC) loss analysis on high-temperature superconductor (HTS) Cross-Conductor (CroCo) cables, in order to evaluate whether they could be utilized for electrical power transmission. The modeling of HTS CroCo cables was based on a cable assembled at the Karlsruhe Institute of Technology (KIT) and the AC loss calculation was based on the H-formulation model implemented in the finite-element method (FEM) software package COMSOL Multiphysics. The AC loss calculations have been carried out for isolated single-phase CroCo cable and three-phase CroCo cables. The AC loss angular dependence of a particular phase of CroCo cables during three-phase operation has been studied. The current distributions of individual tapes within CroCo cables have been investigated.

*Index Terms*— HTS cross-conductor cable, AC loss, power transmission, finite element analysis.

## I. INTRODUCTION

IN recent years, second generation high-temperature superconductor (HTS) rare earth-based tapes have been increasingly tested and used for high-current cables, such as Roebel cable [1], round core cable [2], and twisted stacked-tape cable [3]. Cross-Conductor (CroCo) cable is one of the twisted stacked-tape cables, and is able to offer a high current density, efficient tape usage, good mechanical performance as it can relax stresses from bending and twisting process, and the possibility of fabrication in long lengths [4]. The cable has been initially conceived for high-field magnet applications, but its compact design makes it an attractive possible solution for AC power transmission as well.

There are some previous works on design and test of twisted stacked-tape cables [3, 5], and on CroCo cable [4]. However, an AC study on CroCo cable is missing. This paper presents the modeling and AC loss analysis on isolated single-phase and three-phase CroCo cables. The AC loss angular dependence of a particular phase in three-phase CroCo cables has been investigated. The current distributions of individual tapes within single-phase and three-phase CroCo cables have been studied.

## II. MODELING OF CROCO CABLES

In order to compute the AC loss of CroCo cables with AC current, the *H*-formulation of Maxwell's equations implemented in the finite-element software package COMSOL Multiphysics has been used [6].

The single-phase CroCo cable model investigated here consists of 10 4-mm wide tapes and 22 6-mm wide tapes (32 tapes in total). To achieve better precision, we used the real dimensions of typical superconducting tapes, with a superconducting layer 1 μm thick. The gap between each HTS tape is 0.2 mm. The cable diameter is 9 mm. The three-phase configuration of CroCo cable is shown in Fig. 1. The distance between each phase (cable boundary to boundary) is 5 mm. The power index for the *E-J* power law was 25. An anisotropic B-dependent critical current model [7] was implemented into our modeling:

$$J_c(B) = \frac{J_{c0}}{\left(1 + \frac{\sqrt{(kB_{para})^2 + B_{perp}^2}}{B_c}\right)^b} \quad (1)$$

where $J_{c0} = 4.75 \times 10^{10}$ A/m², $k = 0.25$, $B_c = 0.035$, and $b = 0.6$. A static model to calculate the critical current with equation (1) estimated the total critical current of single-phase CroCo cable 3889 A [7].

As we used power frequency 50 Hz for all the calculation, it was reasonable to neglect the eddy current AC losses in the metal layers [8]. Therefore, for the simulation, the hysteresis losses in the superconducting layer are the only contribution to the AC losses. An overall transport current was imposed into each phase of the cables using the "Global Constraint" in COMSOL, and within each cable the transport current of the cables was let free to distribute among the tapes. The AC loss was calculated by integrating the instantaneous power density on the superconducting domain [9]:

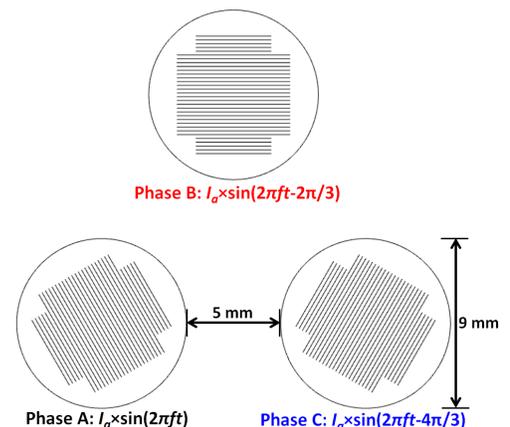

Fig. 1. Configuration of three-phase CroCo cables, with Phase A: $I_a \times \sin(2\pi ft)$, Phase B: $I_a \times \sin(2\pi ft - 2\pi/3)$, Phase C: $I_a \times \sin(2\pi ft - 4\pi/3)$.

B. Shen and T. A. Coombs are with the Electrical Engineering Division, Department of Engineering, University of Cambridge, CB3 0FA, U.K. (e-mail: bs506@cam.ac.uk).

F. Grilli is with the Institute for Technical Physics, Karlsruhe Institute of Technology (KIT), 76131 Karlsruhe, Germany. (e-mail: francesco.grilli@kit.edu).

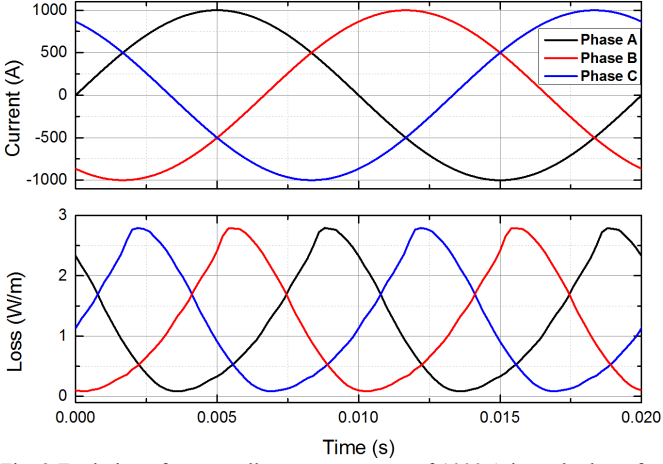

Fig. 2 Evolution of an overall transport current of 1000 A in each phase for three-phase operation (steady state), and the corresponding instantaneous power dissipation (W/m) of each phase.

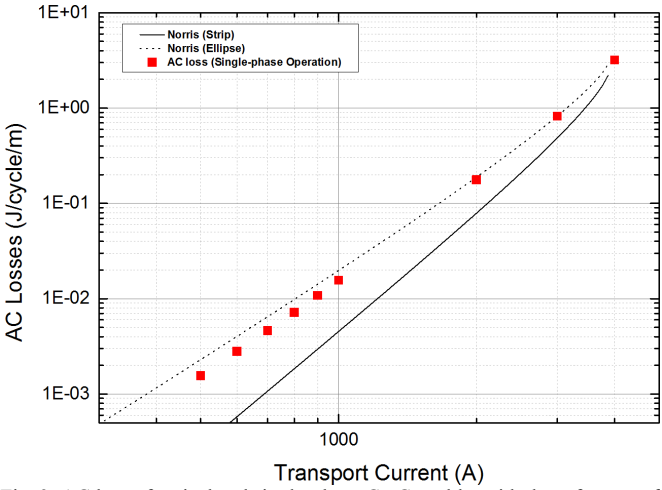

Fig. 3. AC loss of an isolated single-phase CroCo cable, with the reference of Norris's analytical solutions (strip and ellipse).

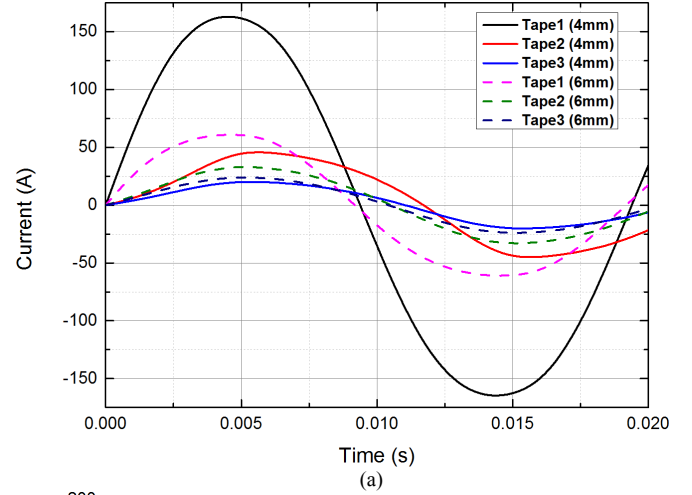

(a)

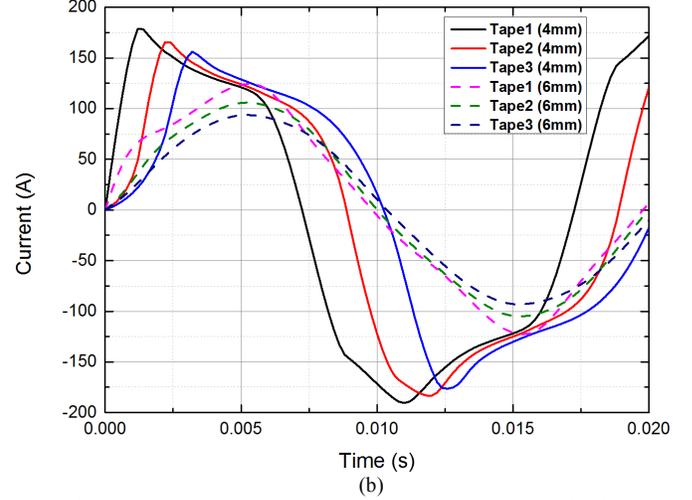

(b)

Fig. 4. Current distributions of several tapes in the top part of CroCo cable during single-phase operation, with overall transport current (a) 1000 A, (b) 3000 A.

$$Q = \frac{2}{T} \int_{0.5T}^{T} \int_{\Omega} \boldsymbol{E} \cdot \boldsymbol{J} \; d\Omega dt \qquad (3)$$

where $\Omega$ is the domain of interest and T is the period of cycle. Fig. 2 presents the evolution of the current and the corresponding instantaneous power dissipation (W/m) of each phase for an overall transport current of 1000 A in each phase during steady-state three-phase operation.

### III. SINGLE-PHASE OPERATION

Fig. 3 presents the AC loss from a CroCo cable during single-phase operation, with the reference of Norris's analytical solutions (strip and ellipse) [10]. The overall transport current was increased from 500 A to 4000 A. For low transport current, the AC loss curve was slightly lower than Norris ellipse. The AC loss curve started to approach and trace the Norris ellipse curve when transport current increased to high value. This is expected because the cross-section of the CroCo is quite similar to that of a round conductor.

Fig. 4 illustrates the current distributions from several individual tapes in the top part of CroCo cable, with overall transport current of 1000 A and 3000 A. As shown in Fig. 4 (a), with transport current of 1000 A, the first 4-mm top tape carry most of the current, whose peak value is around 150 A. Other tapes (2nd and 3rd 4-mm top tapes, and 1st, 2nd, 3rd 6-mm top tapes) carry much less current. In Fig. 4 (b), with transport current of 3000 A, the outermost tapes initially carry most of the current, and only when they get saturated do the inner tapes start carrying some substantial current. Once the outermost tapes get saturated, they start carrying less current: this is because their critical current decreases, due to the self-field generated by total the transport current, which is still increasing (until t=0.005 s). This effect has been observed experimentally in a stacked-tape cable (see Fig. 5 of [11]).

### IV. THREE-PHASE OPERATION

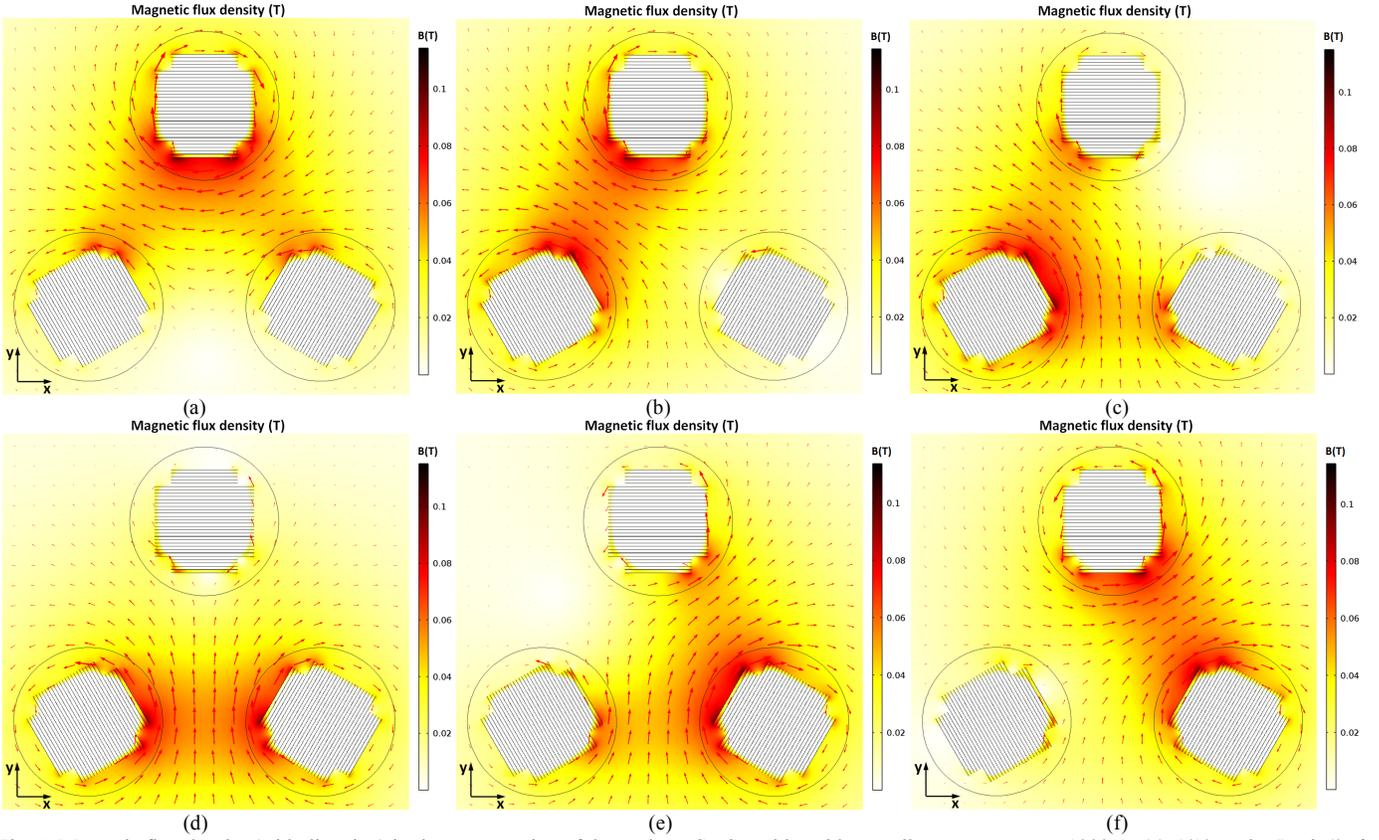

Fig. 5. Magnetic flux density (with direction) in the cross-section of three-phase CroCo cable, with overall transport current 1000 A: (a) 1/12 cycle: $I_a \times \sin(2\pi ft - \pi/6)$, (b) 2/12 cycle: $I_a \times \sin(2\pi ft - 2\pi/6)$, (c) 3/12 cycle: $I_a \times \sin(2\pi ft - 3\pi/6)$, (d) 4/12 cycle: $I_a \times \sin(2\pi ft - 4\pi/6)$, (e) 5/12 cycle: $I_a \times \sin(2\pi ft - 5\pi/6)$, (f) 6/12 cycle: $I_a \times \sin(2\pi ft - 6\pi/6)$.

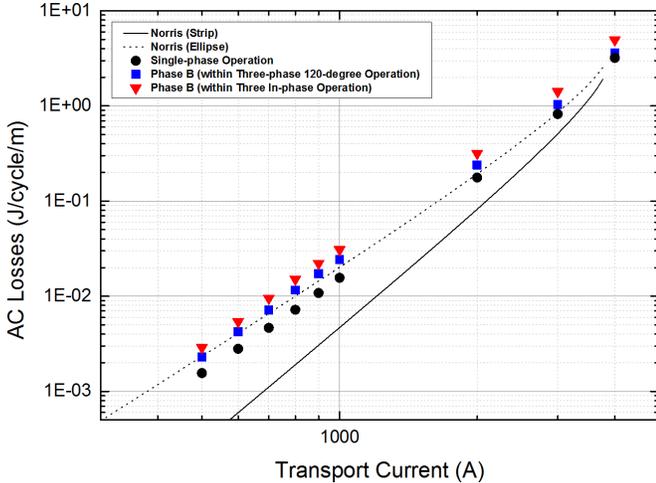

Fig. 6. Comparison: AC loss of a CroCo cable during single-phase operation, AC loss of a CroCo cable during three-phase operation (each phase difference 120°), and AC loss of a CroCo cable during three in-phase operation, with the reference of Norris's analytical solutions (strip and ellipse).

Fig. 5 presents the magnetic flux density (with direction) in the cross-section of three-phase CroCo cable, with overall transport current 1000 A. The current phases are defined in Fig. 1. Herein, we focused on half (1/2) cycle, because it was a circulation of magnetic field variation. Fig. 5 (a), (c) and (e) represent the instants when Phase A, Phase B and Phase C carried the maximum currents, in 1/12 cycle: $I_a \times \sin(2\pi ft - \pi/6)$, 3/12 cycle $I_a \times \sin(2\pi ft - 3\pi/6)$ and 5/12 cycle $I_a \times \sin(2\pi ft - 5\pi/6)$, respectively. Fig. 5 (b), (d) and (f) represents the moments that Phase A, Phase B and Phase C carried the minimum currents, in 2/12 cycle $I_a \times \sin(2\pi ft - 2\pi/6)$, 4/12 cycle $I_a \times \sin(2\pi ft - 4\pi/6)$ and 6/12 cycle $I_a \times \sin(2\pi ft - 6\pi/6)$, respectively. As shown in Fig. 5 (a)-(f), the maximum magnetic flux density in the three-phase operation with overall transport current 1000 A was approximately 0.1 T. Fig. 5 (a)-(f) also depicts the changing orientation (red arrow) of the magnetic field: the magnetic field in the centre of the three-phase CroCo rotates from –x direction (Fig. 5 (a)), and then to +y direction (Fig. 5 (d)), and eventually to from +x direction (Fig. 5 (f)), which was approximately 180 degree.

Fig. 6 shows the comparison of the AC loss of a CroCo cable during single-phase operation, of a CroCo cable during three-phase operation (each phase difference 120°), and – just for reference – of a CroCo cable with three in-phase current, along with Norris's solutions (strip and ellipse). The AC loss curve of the phase with 120 degree phase difference was slightly above Norris ellipse, but always lower than the AC loss curve with three in-phase currents. That was due the stronger magnetic field interaction of three in-phase with respect to the three-phase with 120 degree difference. The AC loss of a CroCo cable during regular three-phase operation with overall transport current 1000 A was approximately 0.024 J/cycle/m, equivalent to 1.2 W/m, which was in the acceptable loss regime for power transmission criteria. However, the AC loss from a HTS CroCo cable during three-phase operation (using 2000 A transport current) is higher than the AC loss from concentric HTS power cables under the same condition [12].

The simulations conducted so far implicitly assume that the three phases maintain their relative orientation (i.e. the whole three-phase cable is either straight or rotates as a whole along its axis). As a further step, we investigated the effect of twisting the individual phases. Fig. 7 illustrates the AC loss angular dependence for the top phase (Phase B) in three-phase CroCo cables, with overall transport current 1000 A. By changing $\theta$ from 0 to 90 degree, the AC loss decreased by approximately 8 %, which implies that the AC loss angular dependence of a particular phase within three-phase CroCo cable is not very significant.

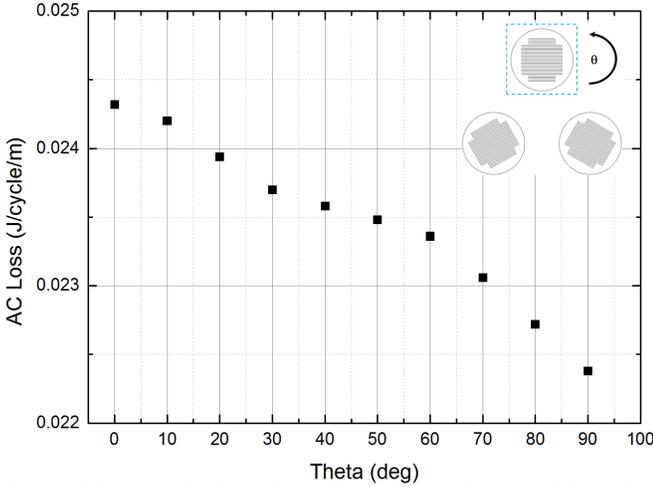

Fig. 7. AC loss angular dependence for top phase (Phase B) of three-phase CroCo cables, with overall transport current 1000 A.

Fig. 8 presents the current distributions of several tapes in the top part of CroCo cable (Phase B), during three-phase operation with overall transport current 1000 A. Similar to the single-phase case shown in Fig. 4 (a), the first top tapes (4 mm tape) carried a large amount of current while others carried much smaller current. Fig. 9 shows the current distributions of several tapes in the bottom part of the CroCo cable (Phase B), with the same conditions as Fig. 8. The bottom Tape 1 (4 mm tape) initially carried most of the current, and this current first increased, saturated, then decreased, which reveals that the current bottom Tape 1 (4 mm tape) carried was around its effective critical current. As shown in both Fig. 8 and Fig. 9, the currents in individual tapes were no longer in phase. This phenomenon was even more obvious in the bottom tapes (Fig. 9).

## V. Conclusion

The modeling and AC loss calculation of isolated single-phase cable and three-phase CroCo cables have been executed by ***H***-formulation model using the FEM package COMSOL Multiphysics. The AC loss curve of an isolated single-phase CroCo cable approximately follows that of Norris's ellipse. The results reveal that the AC loss of one phase CroCo cable during typical three-phase operation is higher than that from isolated single-phase and Norris ellipse, but lower than that during the three in-phase operation. The AC loss from a HTS CroCo cable during general three-phase operation (1000 A transport current per phase) is at an acceptable level for power transmission, but higher than that of HTS cables with concentric design, due to the fact that most of the current flows in the outer tapes of the CroCo. The AC loss angular dependence is not significant when rotating one phase in three-phase CroCo cables. An experimental measurement of the AC losses of the CroCo cable is still necessary. However, the simulation results of this work seem to indicate that, compared to conventional solutions obtained with a concentric tape arrangement, the CroCo cable has too large AC losses and inefficient tape usage. Its use could however become advantageous if a limited cross-section of the cable is a stringent requirement.

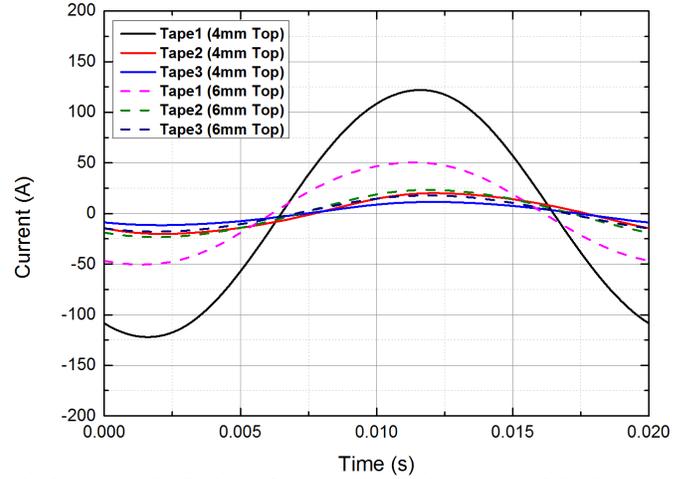

Fig. 8. Current distributions of several tapes in the top part of CroCo cable: during three-phase operation, with overall transport current 1000 A.

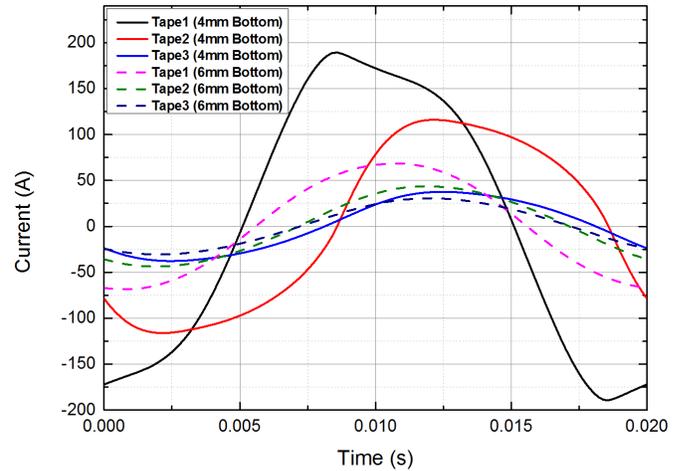

Fig. 9. Current distributions of several tapes in the bottom part of CroCo cable: during three-phase operation, with overall transport current 1000 A.

## VI. Acknowledgments

The authors are grateful to the members of KIT's CroCo team (Walter Fietz, Rainhard Heller, Michael Wolf, Alan Preuss) and to Dustin Kottonau for useful discussions on the CroCo cable. B. Shen would like to acknowledge China Scholarship Council (CSC) for their scholarships and support for overseas students.